\documentclass[prd,article,nofootinbib,twocolumn,preprintnumbers,superscriptaddress,amsmath,amssymb,aps]{revtex4-1}

\usepackage{graphicx,%  Default Latex 2eps package for embedding figures
longtable,%              Useful for long table
color%					Color package
}

\begin{document}

\preprint{IPPP/19/69}
\preprint{MPP/2019/178}

\title{Neutrino mass ordering obscured by non-standard interactions}%

\author{		Francesco Capozzi}
\email{capozzi@mpp.mpg.de}
\affiliation{		Max-Planck-Institut f\"ur Physik (Werner-Heisenberg-Institut), F\"ohringer Ring 6, 80805 M\"unchen, Germany}

\author{		Sabya Sachi Chatterjee}
\email{sabya.s.chatterjee@durham.ac.uk}
\affiliation{		Institute for Particle Physics Phenomenology, Department of Physics, Durham University, Durham, DH1 3LE, UK}

\author{		Antonio Palazzo}
\email{palazzo@ba.infn.it}
\affiliation{ 	Dipartimento Interateneo di Fisica ``Michelangelo Merlin,'' Via Amendola 173, 70126 Bari, Italy}
\affiliation{ 	Istituto Nazionale di Fisica Nucleare, Sezione di Bari, Via Orabona 4, 70126 Bari, Italy}

%\date{{\today}}

\begin{abstract}

One of the major open questions in particle physics is the issue of the neutrino mass ordering (NMO). 
The current data of the two long-baseline experiments NO$\nu$A and T2K, interpreted in the standard 3-flavor scenario,
provide a $\sim2.4\sigma$ indication in favor of the normal neutrino mass ordering. 
We show that such an indication is completely washed out if one assumes the existence of neutral-current non-standard
interactions (NSI) of the flavor changing type involving the $e-\tau$ flavors. This implies that the claim for a discovery
of the NMO will require a careful consideration of the impact of hypothetical  NSI.

\end{abstract}
\maketitle

{\bf {\em Introduction.}}  The neutrino mass ordering (NMO) is one of the most 
important unknown properties in particle physics. Its determination
is crucial for flavor model building and has direct implications for the 
efforts to determine the nature of the neutrino (Dirac  vs Majorana) with neutrinoless double-beta decay searches.
We recall that in the 3-flavor scheme there are three mass eigenstates $\nu_i$ with masses
$m_i\, (i = 1,2,3)$, three mixing angles $\theta_{12},\theta_{13}, \theta_{13}$, and one CP-phase $\delta$.
The mass ordering is defined to be normal (NO)  if $m_3>m_{1,2}$ or inverted  (IO) if  $m_3<m_{1,2}$.
A rich dedicated program of experiments is underway to nailing down the NMO with 
accelerator, atmospheric and reactor neutrinos. 
Quite interestingly the two running long-baseline experiments (LBL)
NO$\nu$A and T2K, in combination with the reactor data
sensitive to $\theta_{13}$, already provide a $\sim2\sigma$ indication in favor of NO~\protect\cite{Note2}.
%%%%%%%%%%%%%%%%%%%%%%%%%%%%%%%%%%%%%%%%%%%%%%%%%%%%%%%
%\footnote{In our present analysis, which makes use of the latest T2K~\cite{T2K_talk} and NO$\nu$A~\cite{Acero:2019ksn} 
%data releases, we find that the indication in favor of NO is now at the $\sim 2.4\sigma$ level.}
%%%%%%%%%%%%%%%%%%%%%%%%%%%%%%%%%%%%%%%%%%%%%%%%%%%%%%%
This trend is  independently corroborated by the Super-Kamiokande atmospheric data~\cite{Abe:2017aap}.
As a consequence the most recent global fits display a preference for NO at the remarkable $\sim3\sigma$ 
level~\cite{Capozzi:2018ubv,deSalas:2017kay,Esteban:2018azc}.

It is important to underline that such and indication is established within the standard 3-flavor scheme, and its
validity in scenarios involving physics beyond the Standard Model is unknown.
In this work we focus on neutral current non-standard neutrino interactions (NSI)
and analyze their impact on the determination of the NMO in NO$\nu$A and T2K.
We find that NSI make the current indication in favor of NO evanescent,
and therefore very fragile. More specifically, we find that the present indication is completely erased in the presence of NSI
of the flavor changing type involving the $e-\tau$ flavors~\cite{Note3}.
%%%%%%%%%%%%%%%%%%%%%%%%%%%%%%%%%%%%%%%%%%%%%%%%%%%%%%%
%\footnote{We stress that the confusion problem we are pointing out is not related to the so-called
%generalized mass ordering degeneracy recently identified in~\cite{Coloma:2016gei} (see also~\cite{Bakhti:2014pva}),
%which arises for the very large values of the NSI couplings (specifically $\epsilon_{ee} \sim -2$) needed to accommodate 
%the ``dark" LMA solution~\cite{Miranda:2004nb} to the solar neutrino problem.}.
%%%%%%%%%%%%%%%%%%%%%%%%%%%%%%%%%%%%%%%%%%%%%%%%%%%%%%%

{\bf {\em Theoretical framework.}} NSI may represent the low-energy manifestation of high-energy physics involving new
heavy states (for a review see~\cite{Farzan:2017xzy,Biggio:2009nt,Ohlsson:2012kf,Miranda:2015dra,Dev:2019anc}) or, 
alternatively, they can be connected to new light mediators~\cite{Farzan:2015doa,Farzan:2015hkd}.
As first recognized in~\cite{Wolfenstein:1977ue}, NSI can modify the dynamics~\cite{Wolfenstein:1977ue,Mikheev:1986gs,Mikheev:1986wj} of the neutrino flavor conversion in matter. 
For prospective studies of NSI at LBL experiments see~\cite{Friedland:2012tq,Coelho:2012bp,Rahman:2015vqa,Coloma:2015kiu,deGouvea:2015ndi,Agarwalla:2016fkh,Liao:2016hsa,Forero:2016cmb,Huitu:2016bmb,Bakhti:2016prn,Masud:2016bvp,Soumya:2016enw,Masud:2016gcl,deGouvea:2016pom,Fukasawa:2016lew,Liao:2016orc,Liao:2016bgf,Blennow:2016etl,Deepthi:2017gxg,Flores:2018kwk,Hyde:2018tqt,Masud:2018pig,Esteban:2019lfo}. 
The NSI can be described by a dimension-six operator~\cite{Wolfenstein:1977ue}
%%%%%%%%%%%%%%%%%%%%%%%%%%%%%%%%%%%%%%%%%%%%
\begin{equation}
\mathcal{L}_{\mathrm{NC-NSI}} \;=\;
-2\sqrt{2}G_F 
%\sum_{\alpha\beta f}
\varepsilon_{\alpha\beta}^{fC}
\bigl(\overline{\nu_\alpha}\gamma^\mu P_L \nu_\beta\bigr)
\bigl(\overline{f}\gamma_\mu P_C f\bigr)
%+ h.c.
\;,
\label{H_NC-NSI}
\end{equation}
%%%%%%%%%%%%%%%%%%%%%%%%%%%%%%%%%%%%%%%%%%%
where $\alpha, \beta = e,\mu,\tau$ refer to the 
neutrino flavor,  $f = e,u,d$ indicates the matter 
fermions, superscript $C=L, R$ denotes the chirality of the 
$ff$ current, and $\varepsilon_{\alpha\beta}^{fC}$ are the strengths 
of the NSI. The hermiticity of the interaction requires
%%%%%%%%%%%%%%%%%%%%%%%%%%%%%%%%%%%%%%%%%%%%%%%%%%%%%%
\begin{equation}
\varepsilon_{\beta\alpha}^{fC} \;=\; (\varepsilon_{\alpha\beta}^{fC})^*
\;.
\end{equation}
%%%%%%%%%%%%%%%%%%%%%%%%%%%%%%%%%%%%%%%%%%%%%%%%%%%%%%
For neutrino propagation in Earth matter, the relevant combinations are
%%%%%%%%%%%%%%%%%%%%%%%%%%%%%%%%%%%%%%%%%%%%%%%%%%%%%%
\begin{equation}
\varepsilon_{\alpha\beta}
\;\equiv\; 
\sum_{f=e,u,d}
\varepsilon_{\alpha\beta}^{f}
\dfrac{N_f}{N_e}
\;\equiv\;
\sum_{f=e,u,d}
\left(
\varepsilon_{\alpha\beta}^{fL}+
\varepsilon_{\alpha\beta}^{fR}
\right)\dfrac{N_f}{N_e}
\;,
\label{epsilondef}
\end{equation}
%%%%%%%%%%%%%%%%%%%%%%%%%%%%%%%%%%%%%%%%%%%%%%%%%%%%%%
where $N_f$ is the number density of fermion $f$.
For the Earth, we can assume neutral and isoscalar matter, implying  $N_n \simeq N_p = N_e$, 
in which case $N_u \simeq N_d \simeq 3N_e$.
Therefore,
%%%%%%%%%%%%%%%%%%%%%%%%%%%%%%%%%%%%%%%%%%%%%%%%%%%%%%
\begin{equation}
\varepsilon_{\alpha\beta}\, \simeq\,
\varepsilon_{\alpha\beta}^{e}
+3\,\varepsilon_{\alpha\beta}^{u}
+3\,\varepsilon_{\alpha\beta}^{d}
\;.
\end{equation}
%%%%%%%%%%%%%%%%%%%%%%%%%%%%%%%%%%%%%%%%%%%%%%%%%%%%%
The NSI affect the effective Hamiltonian for neutrino propagation 
in matter, which in the flavor basis becomes
%%%%%%%%%%%%%%%%%%%%%%%%%%%%%%%%%%%%%%%%%%%%%%%%%%%%%
\begin{equation}
H \;=\; 
U
\begin{bmatrix} 
0 & 0 & 0 \\ 
0 & k_{21}  & 0 \\ 
0 & 0 & k_{31} 
\end{bmatrix}
U^\dagger
+
V_{\mathrm{CC}}
\begin{bmatrix}
1 + \varepsilon_{ee}  & \varepsilon_{e\mu}      & \varepsilon_{e\tau}   \\
\varepsilon_{e\mu}^*  & \varepsilon_{\mu\mu}    & \varepsilon_{\mu\tau} \\
\varepsilon_{e\tau}^* & \varepsilon_{\mu\tau}^* & \varepsilon_{\tau\tau}
\end{bmatrix}\,,
%\right)
\end{equation}
%%%%%%%%%%%%%%%%%%%%%%%%%%%%%%%%%%%%%%%%%%%%%%%%%%%%%%
where $U$ is the Pontecorvo-Maki-Nakagawa-Sakata (PMNS) matrix, which, in its standard parameterization,
depends on three mixing angles ($\theta_{12}, \theta_{13}, \theta_{23}$) and the CP-phase $\delta$.
The quantities $k_{21} \equiv \Delta m^2_{21}/2E$ and $k_{31} \equiv \Delta m^2_{31}/2E$ represent
the solar and atmospheric wavenumbers, having defined $\Delta m^2_{ij} \equiv m^2_i-m^2_j$, while
$V_{\mathrm{CC}}$ is the charged-current matter potential 
%%%%%%%%%%%%%%%%%%%%%%%%%%%%%%%%%%%%%%%%%%%%%%%%%%%%%%
\begin{equation}
V_{\mathrm{CC}} 
\;=\; \sqrt{2}G_F N_e 
\;\simeq\; 7.6\, Y_e \times 10^{-14}
\bigg[\dfrac{\rho}{\mathrm{g/cm^3}}\bigg]\,\mathrm{eV}\,,
\label{matter-V}
\end{equation}
%%%%%%%%%%%%%%%%%%%%%%%%%%%%%%%%%%%%%%%%%%%%%%%%%%%%%%
where $Y_e = N_e/(N_p+N_n) \simeq 0.5$ is the relative electron number density in the Earth crust.
For convenience we introduce the dimensionless quantity $v = V_{\mathrm{CC}}/k_{31}$, which 
gauges the sensitivity to matter effects. Its absolute value
 %%%%%%%%%%%%%%%%%%%%%%%%%%%%%%%%%%%%%%%%%%%%%%%%%%%%%%
\begin{equation}
|v| 
\;=\; \bigg|\frac{V_{\mathrm{CC}}}{k_{31}}\bigg| 
\;\simeq\; 8.8 \times 10^{-2} \bigg[\frac{E}{\mathrm{GeV}}\bigg]\;,
\label{matter-v}
\end{equation}
%%%%%%%%%%%%%%%%%%%%%%%%%%%%%%%%%%%%%%%%%%%%%%%%%%%%%%
will appear in the analytical expressions of the $\nu_\mu \to \nu_e$ transition probability.
We here note that in T2K and NO$\nu$A the first oscillation maximum is achieved respectively 
for $E \simeq 0.6\, {\mathrm{GeV}}$  ($E \simeq1.6\, {\mathrm{GeV}}$).  This implies that matter effects
are a factor of three bigger in NO$\nu$A ($v  \simeq0.14$) than in T2K ($v  \simeq 0.05$). 

In the present work, we consider flavor non-diagonal NSI, that is
$\varepsilon_{\alpha\beta}$'s with $\alpha\ne\beta$.
In particular, we focus  on the couplings $\varepsilon_{e\mu}$
and $\varepsilon_{e\tau}$, which, as will we discuss in detail, lead to a dependency 
on their associated CP-phase in the appearance $\nu_\mu \to \nu_e$ probability~\cite{Note4}.
%%%%%%%%%%%%%%%%%%%%%%%%%%%%%%%%%%%%%%%%%%%%%%%%%%%%%%%
%\footnote{The $\nu_\mu \to \nu_\mu$ disappearance channel is sensitive to the ${\mu-\tau}$ coupling but 
%this can be safely ignored because of the very strong upper bound
%coming from the atmospheric neutrinos $|\epsilon_{\mu\tau}| < 8.0 \times 10^{-3}$~\cite{Aartsen:2017xtt}
%(see also \cite{Mitsuka:2011ty}).}.
%%%%%%%%%%%%%%%%%%%%%%%%%%%%%%%%%%%%%%%%%%%%%%%%%%%%%%% 
We show the results only for $\varepsilon_{e\tau}$, while we will comment about the results obtained for $\varepsilon_{e\mu}$. 
We recall that the current upper bounds (at 90\% C.L.) on the two NSI under consideration are: $|\varepsilon_{e\mu}| \lesssim 0.12$
and $|\varepsilon_{e\tau}| \lesssim 0.36$ as reported in the review~\cite{Farzan:2017xzy}, which refers to the global 
analysis~\cite{Gonzalez-Garcia:2013usa}. These limits are basically corroborated by the more recent analysis~\cite{Esteban:2019lfo}.
Here an important caveat is in order. One should note that the upper bound on $|\varepsilon_{e\tau}|$ found in the global analysis performed in~\cite{Esteban:2019lfo} 
depends on the dataset used for the experiments T2K and NO$\nu$A, which is very sensitive to such parameter. 
In~\cite{Esteban:2019lfo} an older dataset is used compared with that adopted in the present analysis.
As we will see below, the new dataset leads (in IO) to a significant preference for a non-zero value of $|\varepsilon_{e\tau}|$. 
As a consequence its inclusion in the global analysis would imply a sensible relaxation of the upper bound on such a parameter.

Let us consider the transition probability relevant for the LBL experiments T2K and NO$\nu$A.
In the presence of NSI, the probability can be  approximated as 
the sum of three terms~\cite{Kikuchi:2008vq} 
%.........................................................................................................................................
\begin{eqnarray}
\label{eq:Pme_4nu_3_terms}
P_{\mu e}  \simeq  P_{\rm{0}} + P_{\rm {1}}+   P_{\rm {2}}\,,
\end{eqnarray}
%.........................................................................................................................................
which, adopting a notation similar to~\cite{Liao:2016hsa}, take the following expressions~\cite{Note7}
%.........................................................................................................................................
\begin{eqnarray}
\label{eq:P0}
 & P_{\rm {0}} &\,\, \simeq\,  4 s_{13}^2 s^2_{23}  f^2\,,\\
\label{eq:P1}
 & P_{\rm {1}} &\,\,  \simeq\,   8 s_{13} s_{12} c_{12} s_{23} c_{23} \alpha f g \cos({\Delta + \delta})\,,\\
 \label{eq:P2}
 & P_{\rm {2}} &\,\,  \simeq\,  8 s_{13} s_{23} v |\varepsilon|   
 [a f^2 \cos(\delta + \phi) + b f g\cos(\Delta + \delta + \phi)]\,,\nonumber\\
\end{eqnarray}
%........................................................................................................................................
where $\Delta \equiv  \Delta m^2_{31}L/4E$ is the atmospheric oscillating frequency,
$L$ being the baseline and $E$ the neutrino energy, and $\alpha \equiv \Delta m^2_{21}/ \Delta m^2_{31}$.
For compactness, we have used the notation ($s_{ij} \equiv \sin \theta_{ij} $, $c_{ij} \equiv \cos \theta_{ij}$), 
and following~\cite{Barger:2001yr},
we have introduced 
%.........................................................................................................................................
\begin{eqnarray}
\label{eq:S}
f \equiv \frac{\sin [(1-v) \Delta]}{1-v}\,, \qquad  g \equiv \frac{\sin v\Delta}{v}\,.
\end{eqnarray}
%.........................................................................................................................................
In Eq.~(\ref{eq:P2}) we have assumed for the NSI coupling the general complex form
%.........................................................................................................................................
\begin{eqnarray}
\varepsilon = |\varepsilon |  e^{i\phi}\,.
\end{eqnarray}
%.........................................................................................................................................
The expression of $P_2$ is slightly different for  $\varepsilon_{e\mu}$ and  $\varepsilon_{e\tau}$ and,
%.........................................................................................................................................
in Eq. (\ref{eq:P2}), one has to make the substitutions 
%.........................................................................................................................................
\begin{eqnarray}
 \label{eq:P2_NSI}
 a = s^2_{23}, \quad b = c^2_{23} \quad &{\mathrm {if}}& \quad \varepsilon = |\varepsilon_{e\mu}|e^{i{\phi_{e\mu}}}\,,\\
 a =  s_{23}c_{23}, \quad b = -s_{23} c_{23} \quad &{\mathrm {if}}& \quad \varepsilon = |\varepsilon_{e\tau}|e^{i{\phi_{e\tau}}}\,.
\end{eqnarray}
% ............................................................................................................................................
In the expressions given in Eqs.~(\ref{eq:P0})-(\ref{eq:P2}) for $P_0$, $P_1$ and $P_2$, 
the sign of $\Delta$, $\alpha$ and $v$ is positive (negative) for NO (IO).  
Finally, we stress that the expressions of the transition probability provided above are valid for
neutrinos and that the  corresponding expressions for antineutrinos are obtained by flipping in Eqs.~(\ref{eq:P0})-(\ref{eq:P2}) 
the sign of all the CP-phases and of the matter parameter $v$.

From the explicit expression of $P_0$ and $P_1$  one can recognize that their sum 
returns the standard 3-flavor probability while the third term $P_2$  is driven by NSI. 
Now, we can note that the mixing angle $\theta_{13}$, the parameter $v$ and the coupling $|\varepsilon|$ are small and have similar size~\cite{Note6}
%%%%%%%%%%%%%%%%%%%%%%%%%%%%%%%%%%%%%%%%%%%%%%%%%%%
%\footnote{Here we are anticipating that the numerical analysis presented below points toward
%best fit values in the range $|\varepsilon| \sim 0.1-0.4$.}
%%%%%%%%%%%%%%%%%%%%%%%%%%%%%%%%%%%%%%%%%%%%%%%%%%%
$\sim 0.2$, and therefore they can be considered approximately of the same order of magnitude $\mathcal{O}(\epsilon)$, 
while $\alpha \equiv \Delta m^2_{21}/ \Delta m^2_{31} = \pm 0.03$ is $\mathcal{O}(\epsilon^2)$. 
%%%%%%%%%%%%%%%%%%%%%%%%%%%%%%%%%%%%%%%%%%%%%%%%%
%\footnote{Interestingly, a similar decomposition of the conversion probability holds in the presence of 
%a light sterile neutrino~\cite{Klop:2014ima}. In that case, however, the nature of the new interference
%term $P_2$ is kinematical, and it is operative also in vacuum. In fact, the new term is due to 
%the interference of the atmospheric oscillations with those induced by the new large squared-mass 
%splitting related to the sterile state.}
%%%%%%%%%%%%%%%%%%%%%%%%%%%%%%%%%%%%%%%%%%%%%%%%%
We observe that $P_{\rm {0}}$ is positive-definite  and independent of the CP-phases, 
and being $\mathcal{O}(\epsilon^2)$, it provides the leading contribution to the probability. 
The two terms $P_{\rm {1}}$ and $P_{\rm {2}}$ are $\mathcal{O}(\epsilon^3)$ and
are subleading.
In $P_{\rm {1}}$ one recognizes the standard 3-flavor interference term
among the atmospheric and the solar frequencies. The third term $P_{\rm {2}}$ brings the dependency
on the (complex) NSI coupling and it is different from zero only in matter (i.e. if $v \ne 0$). 
This last term is driven by the interference of the matter potential $\varepsilon_{e\tau}V_{CC}$ with the atmospheric wavenumber 
 $k_{31}$ (see the discussion in~\cite{Friedland:2012tq}). 

 {\bf {\em  Data used in the analysis.}}  We extracted the datasets of T2K and NO$\nu$A from the latest data releases 
 provided in~\cite{T2K_talk} and~\cite{Acero:2019ksn}.
 The disappearance channel of T2K consists of 243 $\nu_\mu$-like (140 $\bar{\nu}_\mu$-like) events 
 divided into 28 (19) energy bins. The appearance channel contains three event samples: 75 $\nu_e$-like events and 15 $\bar{\nu}_e$-like events without pions,
 and 15 $\nu_e$-like events with one pion in the final state. The first two samples consist of 23 bins, whereas the last one is divided into 16 bins.
The dataset for the disappearance channel of NO$\nu$A is divided into four quartiles for both the neutrino and antineutrino running mode. 
Each quartile is divided into 19  bins.
In total NO$\nu$A has observed 113 $\nu_\mu$- and 102 $\bar{\nu}_\mu$-like events.  
The dataset for the appearance channel is divided into three samples.
Two of them are based on the particle identification variable (so called low-PID and high-PID)~\cite{Aurisano:2016jvx,Psihas:2019ksa}. 
The third one is the ``peripheral'' sample.  Each of the first two samples is further divided into 6  bins. In
total NO$\nu$A collected 58 $\nu_e$- and 27 $\bar{\nu}_e$-like events. In our analysis we use the software GLoBES~\cite{Huber:2004ka,Huber:2007ji} 
and its additional public tool~\cite{Kopp:NSI}, which can implement NSI.

{\bf {\em  Discussion at the level of the bievents plots.}}  Before presenting the results 
of the analysis we deem it useful to discuss the interplay of
the two experiments T2K and NO$\nu$A at the level of the bievents plots, in which
the $x (y)$ axis reports the number of electron neutrino (antineutrino) events measured
in the experiment under consideration. Such plots are particularly useful as they 
provide a bird eye view of the situation for each experiment and evidence possible 
tensions between different experiments. 
We recall that in the 3-flavor framework the theoretical prediction, for a fixed value of the three mixing
angles $\theta_{12}, \theta_{13}, \theta_{23}$ and of the two squared-mass splittings $\Delta m^2_{31}$
and $\Delta m^2_{21}$, lies on an ellipse which has as running parameter the CP-phase $\delta$ in the range $[0, 2\pi]$.
In the plots presented below we fix the solar parameters~\cite{Note8}
%%%%%%%%%%%%%%%%%%%%%%%%%%%%%%%%%%%%%%%%%%%%%%%%%%%%%%%%%%%
%\footnote{We recall that $\theta_{12}$ and $\Delta m^2_{21}$ are determined by the combination of solar and
%Kamland data.}
%%%%%%%%%%%%%%%%%%%%%%%%%%%%%%%%%%%%%%%%%%%%%%%%%%%%%%%%%%%
$\theta_{12}$ and $\Delta m^2_{21}$ at the best  fit point of the global analysis~\cite{Capozzi:2018ubv},
while the reactor angle $\theta_{13}$ is basically fixed at the best fit value from the reactor experiments~\cite{Note9}.
%%%%%%%%%%%%%%%%%%%%%%%%%%%%%%%%%%%%%%%%%%%%%%%%%%%%%%%%%%%
%\footnote{Technically we treat $\theta_{13}$ as a free parameter taking into account the strong prior coming
% from the reactor experiments (dominated by Daya Bay).}
%%%%%%%%%%%%%%%%%%%%%%%%%%%%%%%%%%%%%%%%%%%%%%%%%%%%%%%%%%%
The values of the atmospheric parameters $\theta_{23}$ and $\Delta m^2_{31}$ are 
taken at the best fit of our own analysis.

%==================================================================
\begin{figure}[t!]
\vspace*{-0.0cm}
\hspace*{-0.2cm}
\includegraphics[height=4.3cm,width=4.3cm]{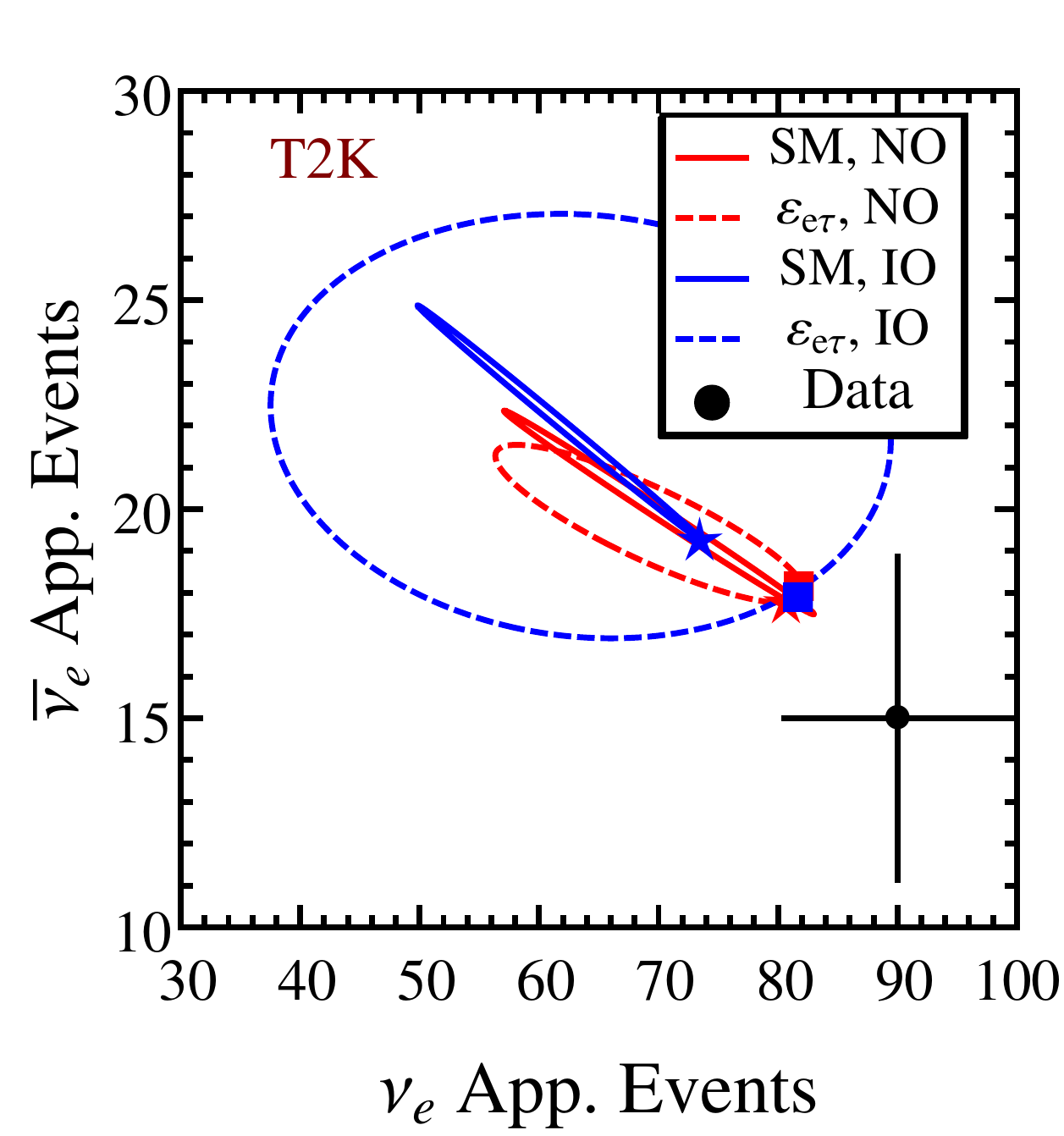}
\includegraphics[height=4.3cm,width=4.3cm]{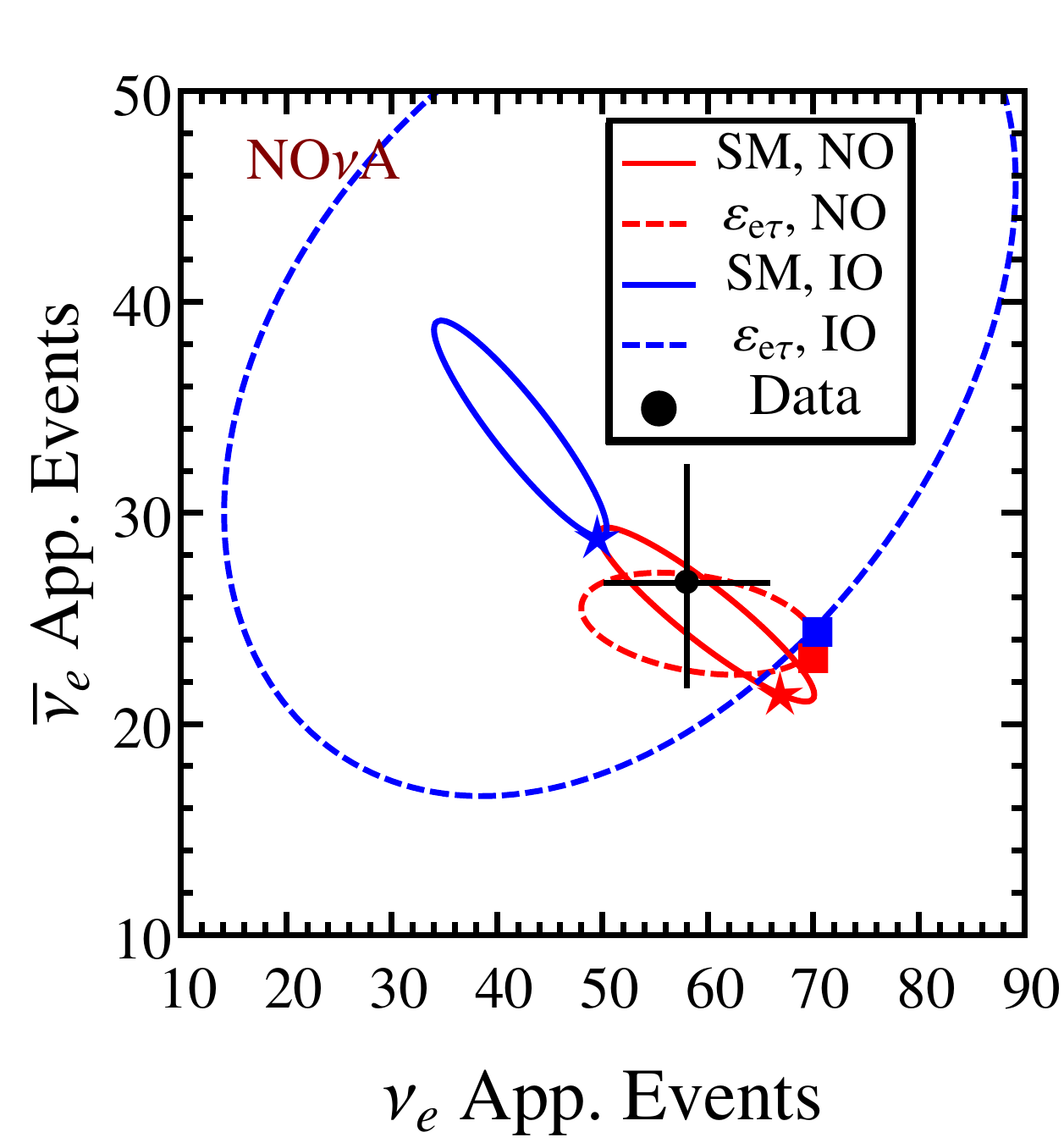}
\vspace*{-0.3cm}
\caption{Bievents plot for the T2K (left panel) and NO$\nu$A setup (right panel). 
The continuos (dashed) ellipse represents the parametric curve corresponding the
SM (SM+NSI) case with running parameter $\delta$ in the range $[0,2\pi]$. The ellipses
and the best fit points located on them are determined by fitting the {\em combination}
of the two experiments.}
\label{Bievent_1}
\end{figure} 
%==================================================================

In Fig.~\ref{Bievent_1} we display the ellipses and best fit points, obtained when one combines {\em together} 
the two experiments T2K and NO$\nu$A. The continuous (dashed) curves correspond to SM (SM + NSI). 
Let us first comment about the SM results. One should note that each of the two experiments constrains the fit of the other one by
forcing the parameters $\delta$, $\theta_{23}$ and $\Delta m^2_{31}$ to assume a common best fit value. 
In particular, in the combination the CP-phase $\delta$ is forced (essentially driven by T2K)
to remain close to $\sim3/2\,\pi$. 
From the left plot we see that in T2K the NO best fit (red star) is closer to the data point with respect to the IO best fit (blue star). 
In the right panel we see that, once the CP-phase $\delta$ is forced to lie close to $\sim 3/2\,\pi$,
in NO$\nu$A there is basically no preference for any of the two hierarchies. 
As a result, in the overall fit,  a moderate global preference for NO is obtained. 
We find $\chi^2_{\mathrm{SM},\, \mathrm{NO}} - \chi^2_{\mathrm{SM},\, \mathrm {IO}} \simeq -5.6$,
corresponding to 2.4$\sigma$  in favor of NO (see also the upper panels of Fig.~\ref{fig:fit}).

In the presence of NSI the additional interference term in the transition probability 
provides much more freedom in the fit. In this case each pair of the NSI parameters $(|\epsilon_{e\tau}|, \phi_{e\tau})$ 
corresponds to a different ellipse in the bievents plot. The amplitude $|\epsilon_{e\tau}|$ 
influences mostly the size of the ellipse, while the phase $\phi_{e\tau}$ determines
the relative length of the two axes as well as the orientation of the ellipse.
The fit of the combination of T2K and NO$\nu$A selects the points (marked as squares)
on those ellipses that provide the best compromise between the two experiments. In NO
the best fit points in the SM+NSI have basically the same distance from the experimental data points
with respect to the SM case. So we expect only a marginal improvement of the fit when
including the NSI. The numerical analysis described in the next section will confirm 
that in NO there is indeed only a $0.7\sigma$ preference for non-zero NSI.
In contrast, for IO in T2K the best fit point is much closer to the experimental point
with respect to the SM case, while in NO$\nu$A the distance between best fit
and data remains basically unchanged.
So in IO we expect a more marked preference for NSI. The numerical analysis
in the next section will confirm that in IO there is indeed a $\sim2.5\sigma$ 
preference for non-zero NSI.
We can also appreciate that in the SM+NSI case the distance of the best fit points (squares)
from the data is basically the same in NO and IO in both experiments, so we expect a similar
goodness of fit of the two mass orderings in the presence of NSI. We find $\chi^2_{\mathrm{SM+NSI},\, \mathrm{NO}} - \chi^2_{\mathrm{SM+NSI},\, \mathrm {IO}} \simeq 0.5$, corresponding to 0.7$\sigma$ in favor of IO (see also the lower panels of Fig.~\ref{fig:fit}).
Therefore, in the presence of NSI the indication in favor of NO found in the standard case gets lost.
 
%==================================================================
\begin{figure}[t!]
\vspace*{-0.0cm}
\hspace*{-0.2cm}
\includegraphics[height=4.3cm,width=4.3cm]{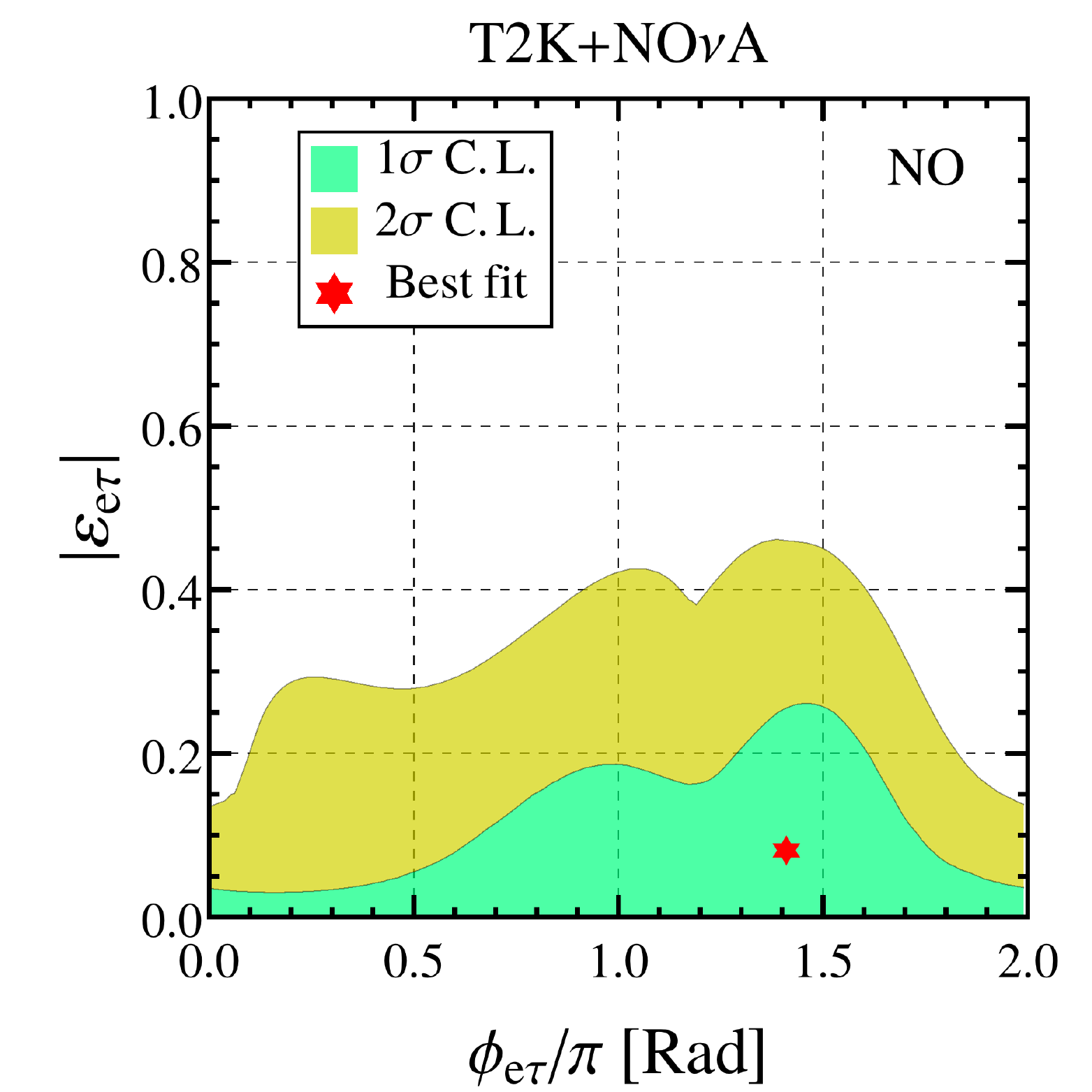}
\includegraphics[height=4.3cm,width=4.3cm]{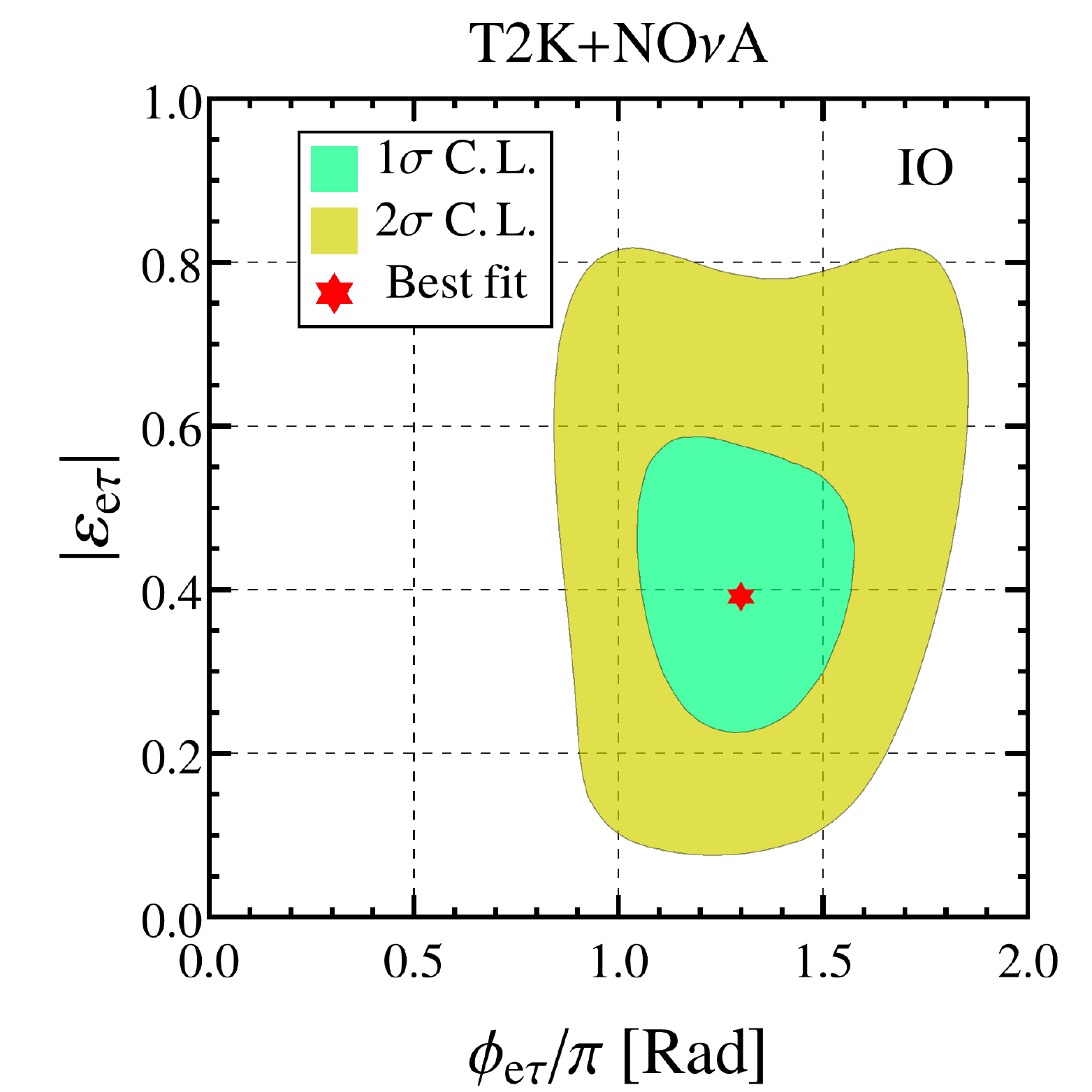}
\vspace*{-0.3cm}
\caption{Allowed regions determined by the combination of T2K and NO$\nu$A for NO (left panel)
and IO (right panel). The contours are drawn at the 1$\sigma$ and 2$\sigma$ level for 1 d.o.f..}
\label{fig:regions}
\end{figure} 
%==================================================================

{\bf {\em Numerical Results.}} Figure~\ref{fig:regions} reports the results of the analysis of the combination of T2K and NO$\nu$A
for NO (left panel) and IO (right panel). Each panel displays the allowed region in the plane spanned by
$|\epsilon_{e\tau}|$ and $\phi_{e\tau}$. The CP-phase $\delta$, the mixing angles $\theta_{23}$ and $\theta_{13}$, and
the squared-mass $\Delta m^2_{31}$ are marginalized away. We show the contours at the
1$\sigma$ and 2$\sigma$ level for 1 d.o.f. and indicate with a star the best fit point.
From the left panel we can appreciate that in NO there is only a weak preference ($\simeq 0.7\sigma$) level for a non-zero 
value of the coupling $|\epsilon_{e\tau}|$, with best fit $|\epsilon_{e\tau}| \simeq 0.09$. In the right panel
we see that the preference for non-zero NSI is stronger, reaching the $2.5\sigma$ significance with
a best fit value $|\epsilon_{e\tau}| \simeq 0.39$. In NO (IO) the CP-phase $\phi_{e\tau}$ has best fit value
$1.42\pi$ ($1.30\pi$).

%==================================================================
\begin{figure}[t!]
\vspace*{-0.0cm}
\hspace*{-0.2cm}
\includegraphics[height=4.3cm,width=4.3cm]{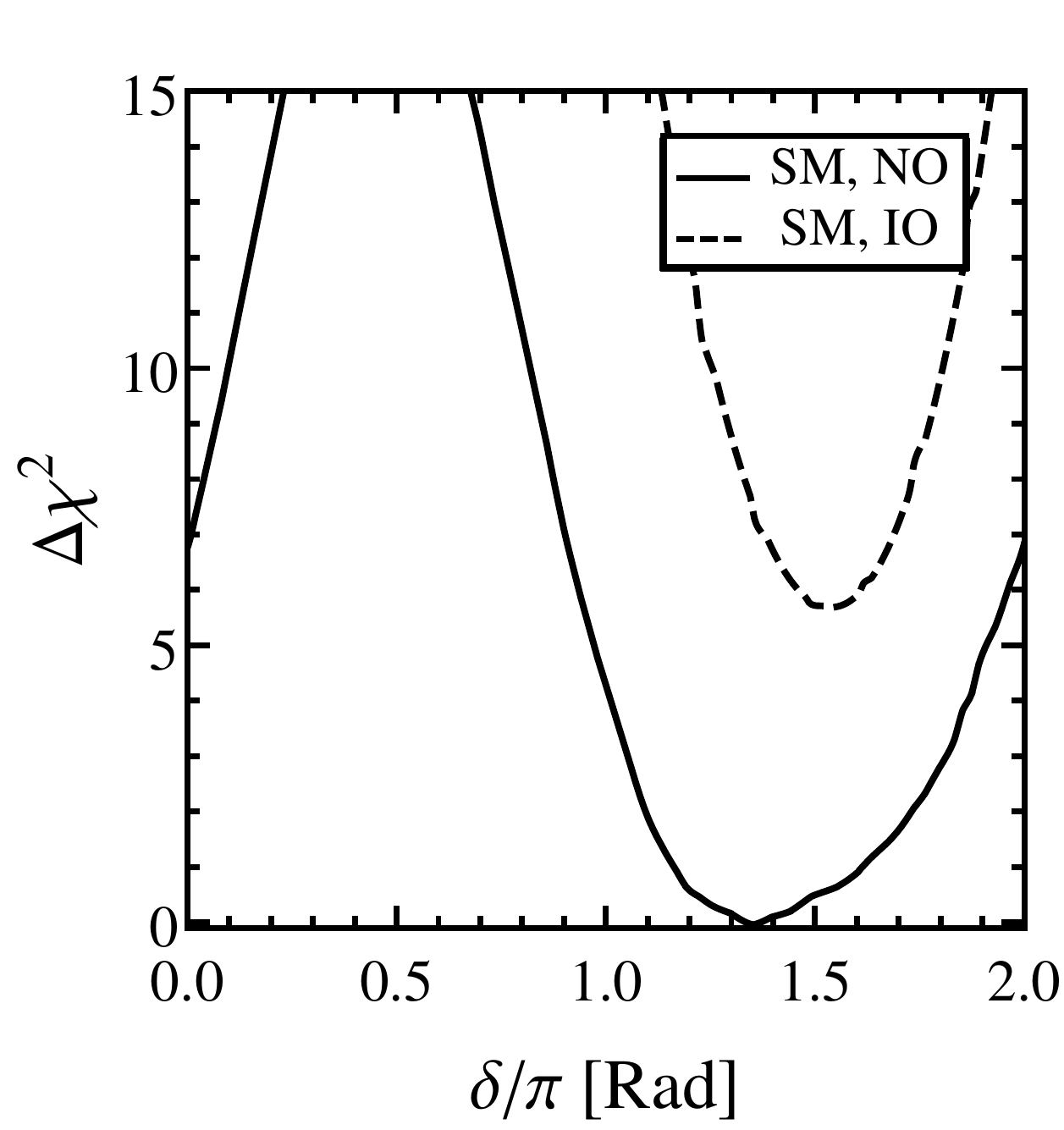}
\includegraphics[height=4.3cm,width=4.3cm]{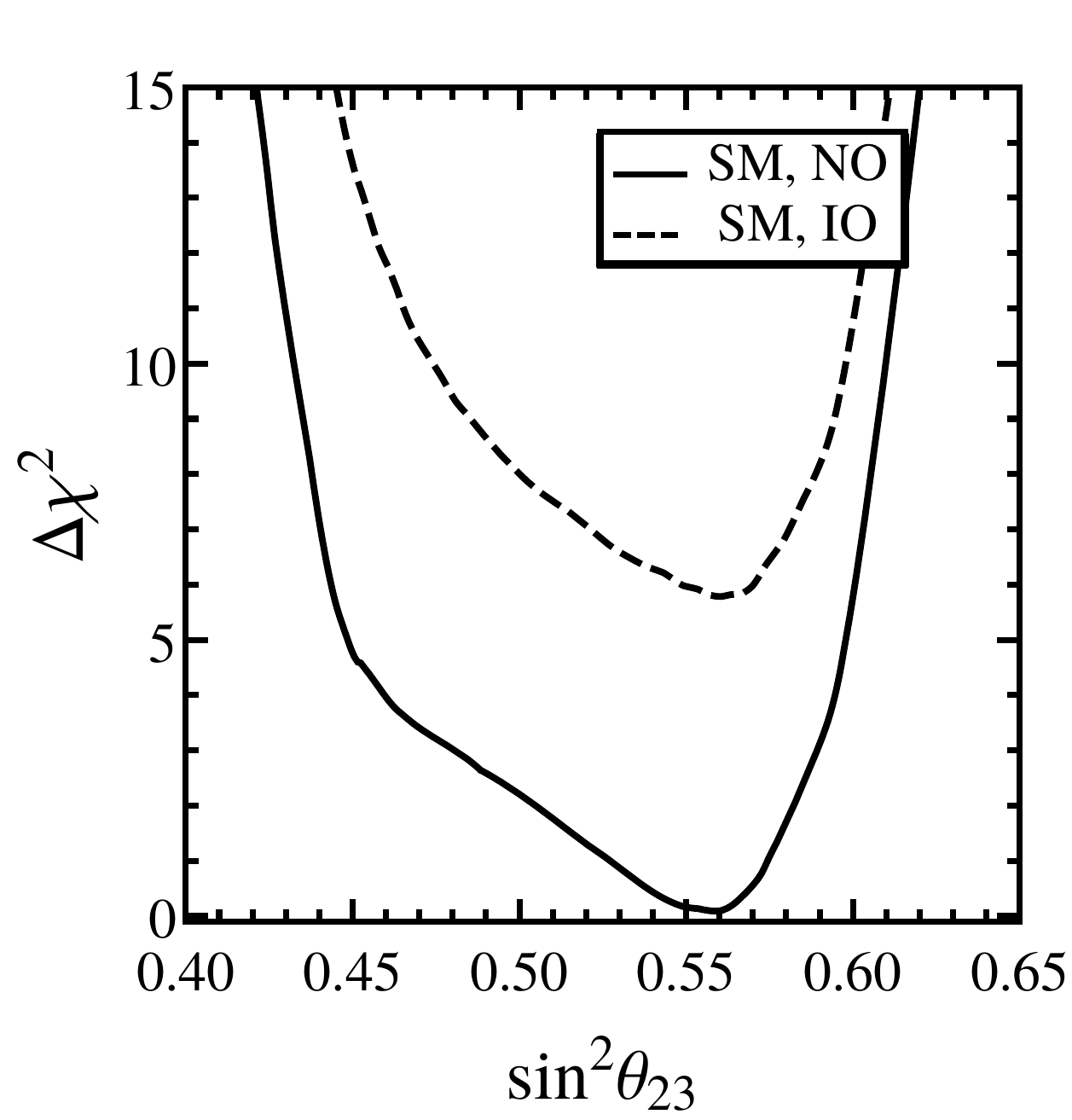}
\includegraphics[height=4.2cm,width=4.2cm]{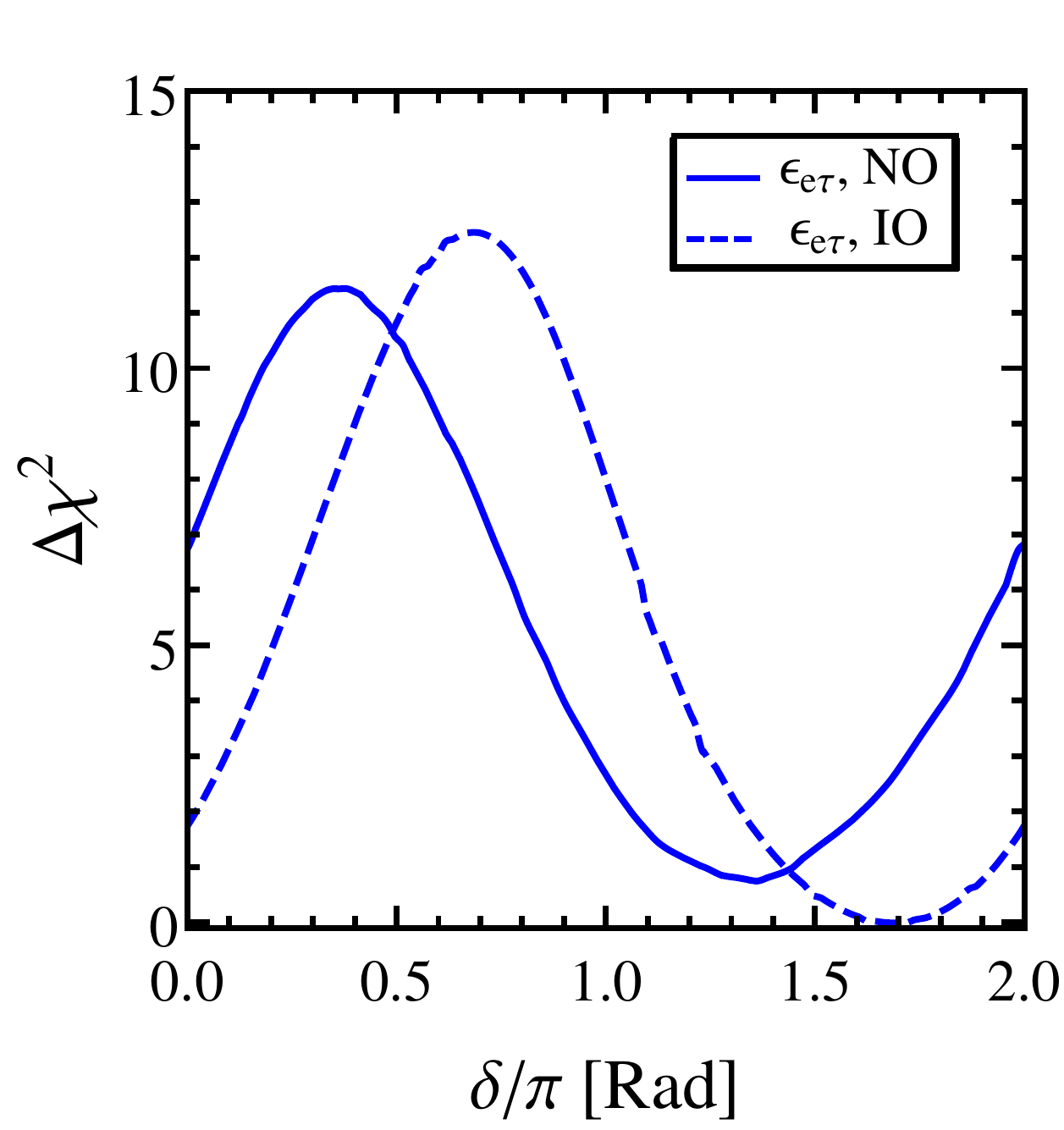}
\includegraphics[height=4.2cm,width=4.2cm]{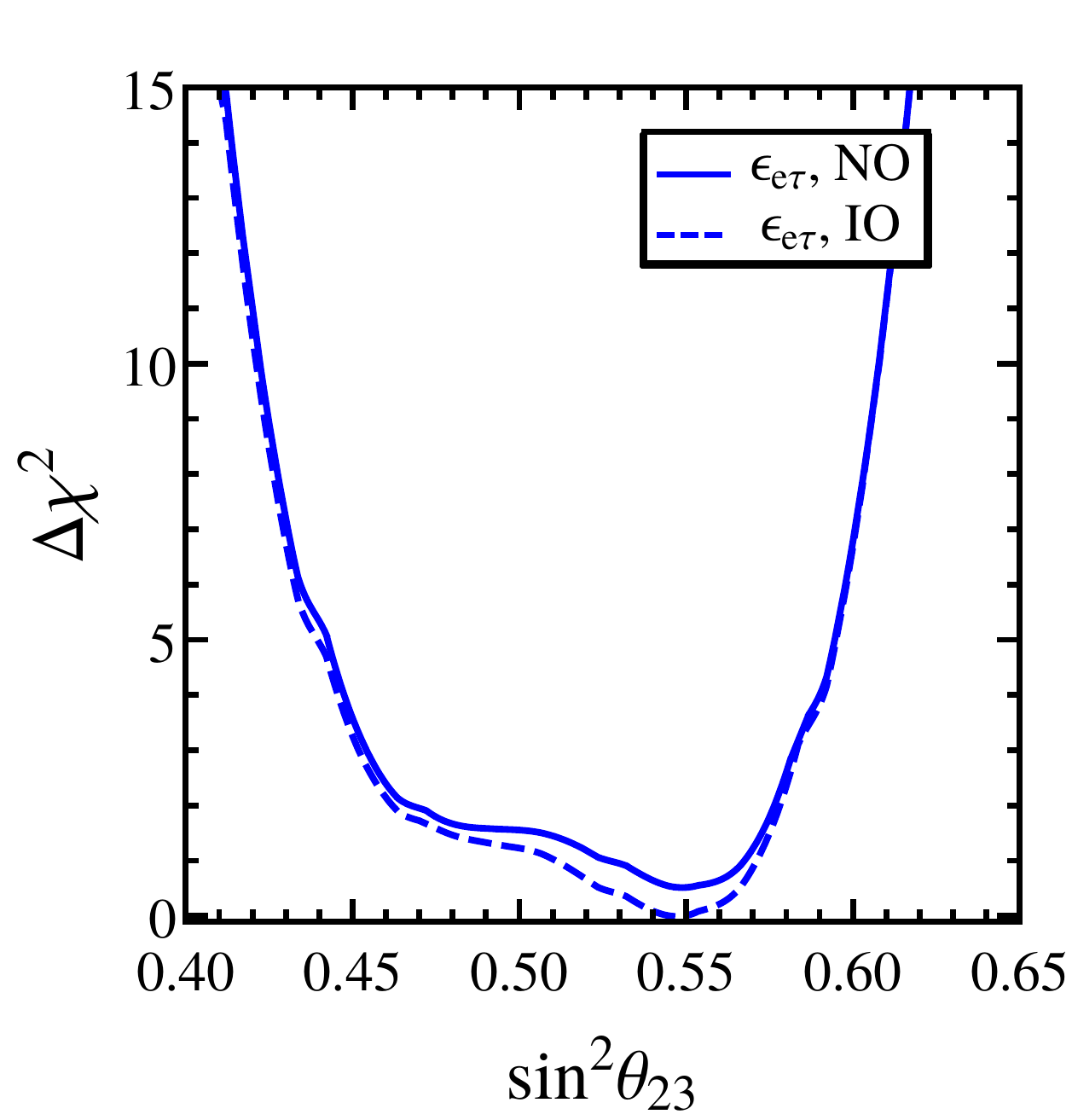}
\vspace*{-0.2cm}
\caption{Estimates of $\delta$ and $\theta_{23}$ for SM (upper panels) and SM+NSI (lower panels) determined 
by the combination of T2K and NO$\nu$A. The continuous (dashed) curves refer to NO (IO).}
\label{fig:fit}
\end{figure} 
%==================================================================

In Fig.~\ref{fig:fit} we show the estimates of the two parameters $\delta$ and $\theta_{23}$.
The two upper (lower) panels report the $\chi^2$ expanded around the minimum value
obtained when the SM (SM+NSI) hypothesis is accepted as true. In each panel we display the results 
obtained by the combination of T2K and NO$\nu$A in NO (continuous lines) and IO (dashed lines).
The undisplayed parameters are marginalized away. We observe that in the SM case there is a 
preference for NO at the $\sim 2.4\sigma$ level. This preference is completely washed out in the presence
of NSI (lower panels), in which case there is even a mild preference for IO (at $\sim 0.7\sigma$).
Therefore, we conclude that the current indication in favor of NO is fragile with respect to the perturbations
induced by the NSI under study.  From the left panels we can observe that the reconstruction of $\delta$ deteriorates
in the presence of NSI. Notably, the values of $\delta$ close to $\pi/2$ are 
rejected at a much lower statistical level with respect to the SM case
($\sim3\sigma$ instead of $\sim5\sigma$). In addition, in the presence of NSI  the CP conserving values
$\delta = (0,\pi)$ are rejected at a lower confidence level.
Concerning $\theta_{23}$ we see that
in the SM case (right upper panel) there is a moderate preference for non-maximal mixing ($\Delta \chi^2 \simeq 2.1$)
and for the higher octant, the symmetric value of the best fit point in the lower octant ($\sin^2 \theta_{23} \simeq 0.45$)
being markedly disfavored ($\Delta \chi^2 \simeq 4.5$). In the presence of NSI 
both the preference for a non-maximal mixing and that for the higher octant found in the SM sensibly decrease.
In this case maximal mixing is disfavored only at $\Delta \chi^2 \simeq 1.3$  and the lower octant
at $\Delta \chi^2 \simeq 2.9$.

It is worthwhile to make some comments on the results of the analysis
obtained considering the interaction in the $e-\mu$ sector. Differently from the $e-\tau$ sector
we find a significant preference for non-zero values of $|\varepsilon_{e\mu}|$ both in NO and IO.
We obtain best fit value $|\varepsilon_{e\mu}|=0.15$ ($|\varepsilon_{e\mu}|=0.10$) for NO (IO)
with statistical significance of $1.6\sigma$ ($1.5\sigma$). The corresponding best fit value
of the CP-phase $\phi_{e\mu}$ is $1.78\pi$ ($1.82\pi$) for NO (IO). Since there is approximately
the same improvement of the fit  in NO and IO, the indication in favor of NO remains almost
unaltered with respect to the SM case. In fact, we find that NO is preferred over IO at
the $2.5\sigma$ level similar to the SM case. Therefore, the fragility of the indication in favor of the NO appears
only when considering the non-standard interactions in the $e-\tau$ sector.

In this letter we have focused on the real data provided by NO$\nu$A and T2K.
Needless to say, it is interesting to consider the impact of the degeneracy problem
we have found in real data on the future LBL experiments, which are expected to have
a higher sensitivity to the NMO. In the Supplemental Material (which includes Refs.~\cite{Acciarri:2015uup,Abe:2015zbg,Abe:2016ero,Baussan:2013zcy,Alion:2016uaj,Huber:2004ka,Huber:2007ji,Kopp:NSI}),
we show explicitly what may be expected from DUNE.  Here, we limit ourselves to 
summarize the basic result, by stating that the degeneracy issue under consideration can be
 resolved by DUNE only if $\delta$ will be confirmed to lie sufficiently 
close to its current best fit value $\delta \sim 1.5\pi$. Also we have checked
that ESS$\nu$SB, T2HK and T2HKK will have less chances to resolve
the issue with respect to DUNE. 

{\bf {\em Conclusions.}} In this letter we have investigated the impact of NSI on
the interpretation of T2K and NO$\nu$A data. Our main result is that the indication in favor
of NO valid in the standard 3-flavor scheme does not hold anymore if one assumes the existence of neutral-current
non-standard interactions (NSI) of the flavor changing type involving the
$e-\tau$ ($\varepsilon_{e\tau}$) flavors.  We have also investigated the potential
role of the future LBL experiment DUNE in resolving the degeneracy issue that plagues
the present data. An unambiguous  determination of the NMO in the presence of NSI
will be possible only if the CP-phase $\delta$ will be confirmed
to lie not too far from the present best fit value around $\delta \sim 1.5\pi$. 
We conclude this letter by underlining that, apart from the LBL accelerator data, it would be very interesting to complement 
our study considering the existing and future 
atmospheric neutrino data, which may help to resolve the NMO ambiguity.
Finally, our work clearly evidences the importance of investigating the NMO exploiting 
different types of experiments, such as JUNO, which are less sensitive to
(standard and non-standard) matter effects.
\vspace{0.5cm}

\begin{acknowledgments}

\noindent The work of F.C. is supported partially by the Deutsche Forschungsgemeinschaft through Grant No. EXC 153 (Excellence Cluster ``Universe") and Grant No. SFB 1258 (Collaborative Research Center ``Neutrinos, Dark Matter, Messengers") as well as by the European Union through Grant No. H2020-MSCA-ITN-2015/674896 (Innovative Training Network ``Elusives''. S.S.C. acknowledges funding support from the European Union's Horizon 2020 research and innovation programme under the Marie Sklodowska-Curie grant agreement No. 690575 during his InvisiblesPlus secondment in Fermilab where substantial part of this work was completed. A.P. acknowledges partial support by the research grant number 2017W4HA7S ``NAT-NET: Neutrino and Astroparticle Theory Network'' under the program PRIN 2017 funded by the Italian Ministero dell'Istruzione, dell'Universit\`a e della Ricerca (MIUR) and by the research project {\em TAsP} funded by the Instituto Nazionale di Fisica Nucleare (INFN). 

\end{acknowledgments}

\bibliographystyle{h-physrev41}
\bibliography{NSI-References}

\clearpage
\newpage
\maketitle
\onecolumngrid
\begin{center}
\textbf{\large Supplemental Material} \\ 
\vspace{0.05in}
%{ \it \large Supplemental Material}\\ 
%\vspace{0.05in}

\end{center}
%%%%%%%%%% Merge with supplemental materials %%%%%%%%%%
%%%%%%%%%% Merge with supplemental materials %%%%%%%%%%
\setcounter{equation}{0}
\setcounter{figure}{0}
\setcounter{table}{0}
\setcounter{section}{1}
\renewcommand{\theequation}{S\arabic{equation}}
\renewcommand{\thefigure}{S\arabic{figure}}
\renewcommand{\thetable}{S\arabic{table}}
\newcommand\ptwiddle[1]{\mathord{\mathop{#1}\limits^{\scriptscriptstyle(\sim)}}}

Here we consider the future implications of the degeneracy problem
we have found in the current data of NO$\nu$A and T2K. New LBL facilities (DUNE~\cite{Acciarri:2015uup}, T2HK~\cite{Abe:2015zbg}, 
T2HKK~\cite{Abe:2016ero}, and ESS$\nu$SB~\cite{Baussan:2013zcy}) are 
under consideration, which are expected to improve the identification of the NMO.
With this purpose, we consider DUNE as a case study, since such an experiment
is the most sensitive to the NMO and to the NSI. This occurs because the impact of 
matter effects is larger in this setup compared with the other ones.
In addition, making use of a broad-band neutrino flux, DUNE is expected to acquire some
extra sensitivity from the shape information of the energy spectrum. To simulate the DUNE setup,
we consider the reference design as mentioned in the Conceptual Design Report~\cite{Acciarri:2015uup},
and use the necessary simulation files~\cite{Alion:2016uaj} for the GLoBES software~\cite{Huber:2004ka,Huber:2007ji} 
and its new tool~\cite{Kopp:NSI}, which can include the effect of NSI.
We have marginalized over $\theta_{13}$ in DUNE simulation with the same prior taken for the real data analysis.
We have also marginalized over $\Delta m^2_{31}$ in the test, where we have taken the true value
equal to $2.49\times 10^{-3}$ eV$^2$ obtained from real data analysis. Solar parameters have been kept fixed.

%==================================================================
\begin{figure}[h!]
\vspace*{-0.0cm}
\hspace*{-0.2cm}
\includegraphics[height=7.3cm,width=7.3cm]{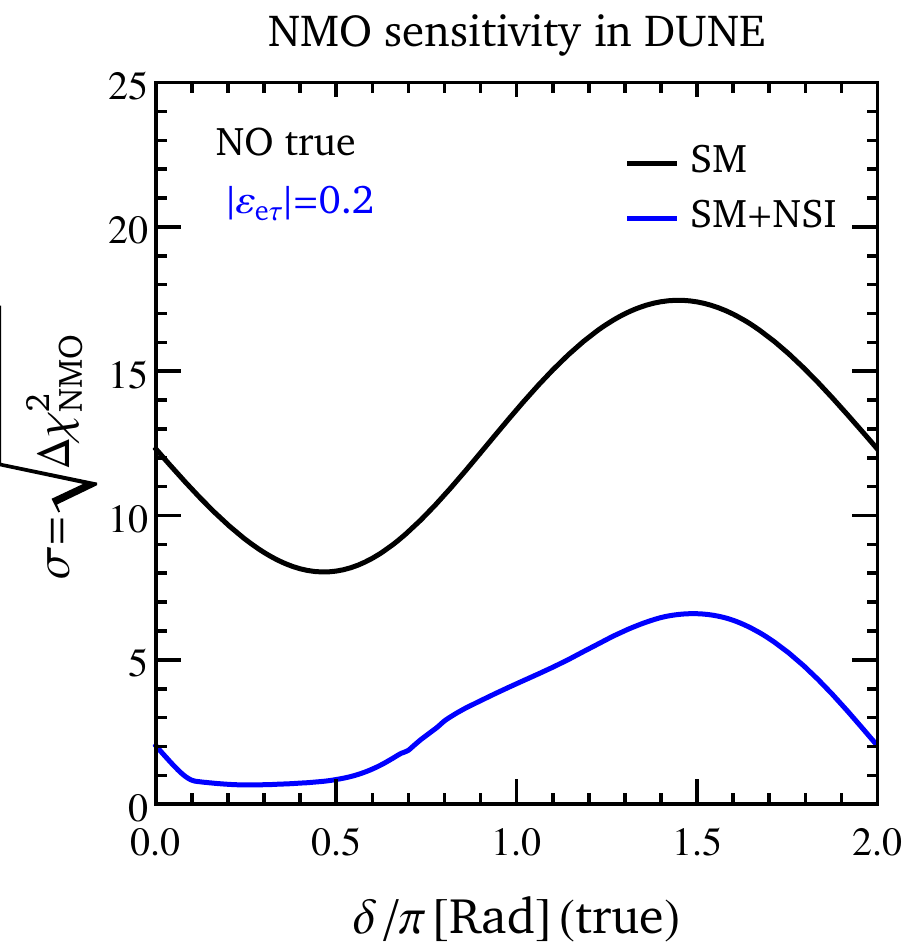}
\vspace*{-0.2cm}
\caption{Sensitivity of DUNE to the NMO. The black curve represents the standard 3-flavor case.
The blue curve is obtained in the presence of NSI.}
\label{fig:DUNE}
\end{figure} 
%==================================================================

In Fig.~\ref{fig:DUNE}, we show the discovery potential of DUNE of the NMO
as a function of $\delta$ assuming the NO as the true case.  The black curve represents the sensitivity in the 
standard 3-flavor scheme. It is very high, varying between a minimum of 8$\sigma$
(reached for $\delta\sim 0.5\pi$) and a maximum of 17$\sigma$ (assumed for $\delta\sim 1.5\pi$).
The blue curve is obtained adding a NSI in the $e-\tau$ sector. More specifically,
we have taken the true value $|\varepsilon_{e\tau}| = 0.2$, which is intermediate between
the values we find as best fits in the analysis of T2K and NO$\nu$A data,
respectively for NO and IO. We have marginalized over the test value of  $|\varepsilon_{e\tau}|$
assuming a gaussian prior with uncertainty of 20$\%$. We have fixed the true value of  $\theta_{23}$ ($\sin^2\theta_{23} = 0.55$)
and marginalized over its test value. We have marginalized over the test value of $\delta$ and over 
both the true and test values of the CP-phase $\phi_{e\tau}$. The blue curve shows a drastic 
deterioration of the sensitivity for all values of $\delta$. In the range of $[0, 0.7\pi]$, the sensitivity 
drops below the $2\sigma$ level. Around $\delta \sim 1.5\pi$, which is  close to the value preferred
by current data, the sensitivity reaches the maximum value of $\sim6.5\sigma$ level. 
Hence, we can state that the degeneracy issue we have found can be resolved by DUNE
only if $\delta$ will be confirmed to lie sufficiently close to its current best fit value $\delta \sim 1.5\pi$.
For completeness, we have performed analogous simulations for T2HK and T2HKK. In both
cases, we find that the sensitivity to NMO in the presence of NSI is very low and below
the 2$\sigma$ level for almost all values of $\delta$. Therefore, we conclude that,
among the future LBL experiments, DUNE will have the best chances to remove
the NMO confusion issue plaguing the present data. Finally, we mention that if
no prior on  the test value of $|\varepsilon_{e\tau}|$ is assumed,  the sensitivity
of DUNE to the NMO further deteriorates, never reaching the $3\sigma$ level.

\end{document}